# Characterizing Peace Through Scientific Keywords

Gita Ghiasi[*], Rabeeh Parhizkari[**] and Tanja Tajmel[***]
[*]gghiasi@uottawa.ca; **Rparh078@uottawa.ca;
*School of Engineering Design and Teaching Innovation, University of Ottawa, Canada*
[***] tanja.tajmel@concordia.ca
*Center for Engineering in Society, Concordia University, Canada*

## Abstract

This study aims to show that peace, as reflected in scientific research, is not a neutral or uniform concept but is constructed differently across disciplines and geopolitical contexts. To do this, this study analyzes differences in the use of keywords among scholarly publications aligned with SDG 16 across various disciplinary, societal, and national contexts, and identifies which keywords are most strongly associated with countries exhibiting higher versus lower levels of societal peace. Results show that while peace-related research is strongly centered on violence, it is interpreted through distinct lenses: technological and forensic approaches in biomedical and NSE fields, and institutional, legal, and societal perspectives in the social sciences. At the country level, violence remains a central anchor, but its framing varies with levels of peace, with higher-peace countries embedding it within welfare, health, and social systems, and lower-peace countries emphasizing conflict, security, and geopolitics. These results could contribute to shaping both research priorities and science policy initiatives.

## 1. Introduction

Peace has become a central topic of discussion nowadays. However, media representations often frame peace primarily as the absence of war or conflict rather than as the advancement of human rights. While much of the existing literature has focused on conflict prevention, there is a growing scholarly interest in understanding the conditions that promote sustained peace (Coleman et al., 2021). Highly peaceful societies are not simply defined by the absence of conflict or violence; rather, they are characterized by non-warring norms, values, positive social interconnectedness, and peace leadership (Fry et al., 2021). This shift in perspective has led to greater emphasis on the concept of positive-peace (Diehl, 2016), which seeks to identify and examine the active social forces that support peace in societies (Liebovitch et al., 2023). This is to the extent that the language of “sustaining peace” is introduced as a counterpoint to “peacebuilding”, aiming to redefine peace as an independent concept, detached from its association with conflict (Mahmoud & Makoond, 2017). This language reflects the mutual reinforcement between sustaining peace and the UN’s sustainable development goals (SDGs), where each supports the achievement of the other (Mahmoud & Makoond, 2017). Along these lines, the discourse of peace in communication has emerged as a critical area of study, emphasizing language as a powerful tool that shapes perceptions of reality and influences human interaction, including conflicts. Empirical findings suggest that higher-peace societies tend to exhibit media narratives centered around themes such as finance, daily activities, and health, whereas lower-peace societies are more frequently associated with topics related to politics, government, and legal issues (Prasad et al., 2025). Another similar study on online news media sources suggested that lower-peace countries are characterized by words of government, order, control, and fear (e.g., government, state, court), while higher-peace countries are characterized by words of optimism for the future and fun (e.g., time, like, game) (Liebovitch et al., 2023). As a result, linguistic patterns and peace narratives in news media and communication not only could reinforce disparities in how peace is defined in a society but also actively contribute to what is perceived as the “peacefulness” of a society.Scholarly publications represent an important channel for scholarly communication through which knowledge is disseminated, and the language used within these publications can shape how concepts such as peace or peacefulness are defined for different societies. Over the past decade, numerous studies have attempted to map the SDGs within scientific research—an approach commonly referred to as SDG mapping (Armitage et al., 2020)—to assess the extent to which scientific outputs align with the SDGs. However, relatively limited attention has been given specifically to SDG 16 (Peace, Justice, and Strong Institutions) (Ghiasi et al., 2021), which encompasses targets including reducing violence and related deaths; ending abuse, exploitation, trafficking, and violence against children; promoting the rule of law and equal access to justice; combating organized crime and illicit financial and arms flows; reducing corruption and bribery; developing effective, accountable, and transparent institutions; ensuring inclusive and representative decision-making; strengthening the participation of developing countries in global governance; providing universal legal identity; ensuring public access to information and protecting fundamental freedoms; strengthening national institutions to prevent violence and combat terrorism and crime; and promoting non-discriminatory laws and policies (United Nations, 2015). Departing from conventional SDG 16 mapping approaches, this study analyzes differences in the use of keywords among scholarly publications aligned with SDG 16 across various societal and national contexts, and identifies which keywords are most strongly associated with countries exhibiting higher versus lower levels of societal peace.

## 2. Methods

This study extracted SDG 16–aligned publications (hereafter referred to as peace-related papers) from the Web of Science database covering the period from 1995 to 2025. Although the SDGs were formally introduced in 2015, the concepts of peace, justice and institutions had been extensively studied prior to their adoption. In the Web of Science Core Collection, publications are assigned to SDGs using a combination of automated citation network analysis (Citation Topics) and expert curation by the Institute for Scientific Information (ISI) (Garcia, 2022). This approach is different from other major databases (such as Scopus or Dimensions) as it clusters publications based on citation networks and includes both algorithmic and manual curation to minimize errors and false positives. Moreover, since it relies on citation networks rather than SDG-related English keywords, this approach could also capture relevant research that does not explicitly include those keywords.

In total, 219,996 peace–related articles were identified, of which 113,690 included author-provided keywords. As this study aims to capture the language used by scholars, Keyword Plus terms were excluded from the analysis. Unlike author keywords, Keyword Plus is algorithmically generated based on frequently occurring words or phrases in the titles of cited references, rather than reflecting the authors' own terminology. Disciplines are assigned to papers based on the National Science Foundation's (NSF) journal field classifications. As noted earlier, the objective of this study is not to conduct SDG 16 mapping, but rather to identify which keywords co-occur more frequently across different national contexts. National contexts are defined by whether a nation or a nationality name appears either in the title or in the author keywords. To achieve this, a series of data preprocessing steps was applied to standardize and refine the keyword dataset. First, country names and nationalities were cleaned and unified by replacing nationality terms with their corresponding country names. This approach ensured consistency, as references to a country—whether through nationality or country name—were treated equivalently. Second, compound keywords were tokenized by splitting phrases into individual words, allowing for more granular keyword frequency analysis. For example, the phrase "Ukrainian people" was separated into two distinct tokens: "Ukraine" and "people". This enabled more precise identification of commonly associated terms across countries. Finally, all keywords were normalized to their root forms to address lexical variation. Specifically, natural language processing (NLP) techniques, including Porter stemming and WordNet-based lemmatization, were employed to consolidate semantically similar terms. Following data cleaning and preprocessing, the analysis focused on identifying the most frequently occurring keywords associated with each country. Countries are selected based on their 2026 Global Peace Index (GPI) rank by The Institute for Economics & Peace (Institute for Economics & Peace, 2026) , and if the number of papers *exceeds 100*. Therefore, if the number of identified papers is less than 100, the country is omitted from the analysis.

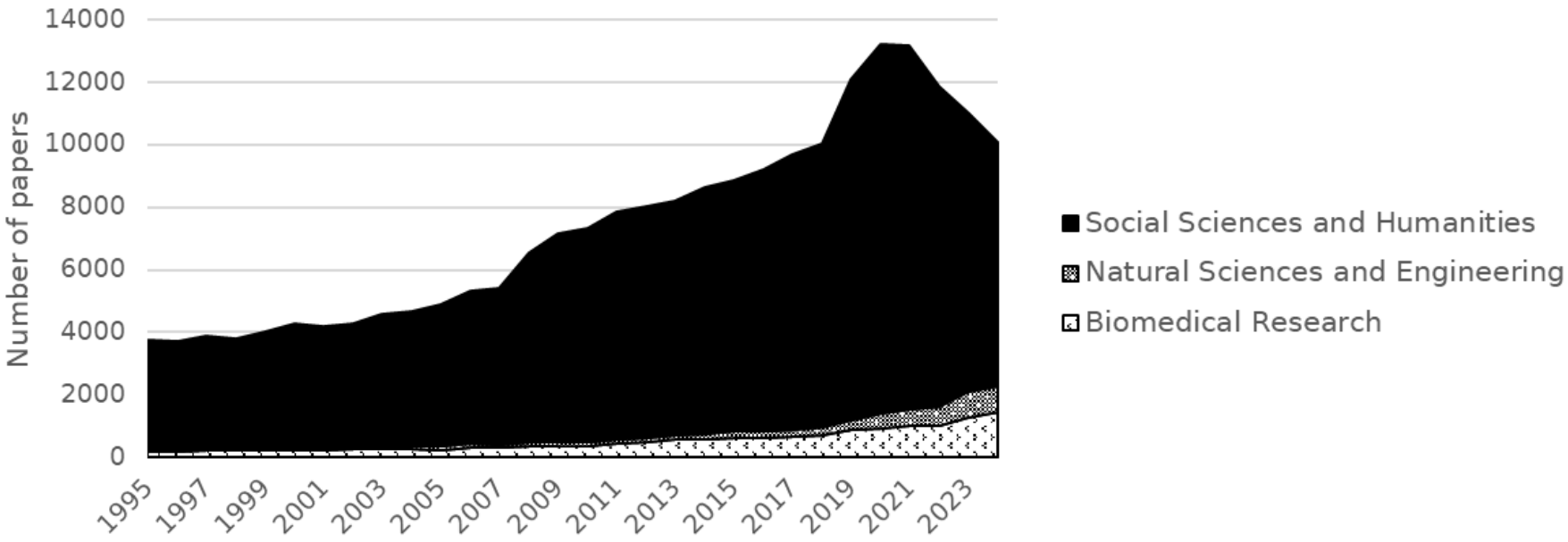

Figure 1- Number of SDG-16 related papers by discipline

## 3. Results

Our analyses (Fig. 2) suggest that the top keywords in peace-related papers in biomedical research may follow a primary theme of sexual, domestic, and intimate partner violence with forensic and health dimensions. This topic appears to capture research on gender-based violence, with a strong emphasis on sexual assault, domestic abuse, and intimate partner violence. The prominence of terms such as "forensic," "analysis," and "identification" may indicate a significant focus on investigative and evidentiary approaches, while the inclusion of "health" may reflect the physical and psychological consequences of assault or violence. The presence of "women," "gender," and "adolescents" may further suggest a demographic emphasis on populations that are disproportionately affected by such forms of violence.

In NSE, this cluster may reflect a technologically driven research stream that focuses on the development and application of computational tools for forensic detection and analysis. It may highlight the increasing integration of artificial intelligence, advanced imaging, and data analytics into crime investigation processes. The presence of "spectroscopy" may suggest the use of advanced analytical techniques for the chemical characterization of forensic evidence, often integrated with imaging and machine-learning approaches to help with detection and identification.

In the social sciences and humanities, while comprising the largest proportion of papers, this cluster may capture the social and political dimensions of gender-based violence, with particular emphasis on sexual, domestic, and intimate partner violence. The presence of terms such as "policy," "international," and "justice" may indicate the role of legal frameworks and governance systems in addressing violence, while "gender," "women," and "theory" may reflect a strong grounding in gender analysis and social theory. The inclusion of "crime" and "police" may further suggest a focus on how violence is conceptualized, regulated, and addressed within broader societal and political contexts.

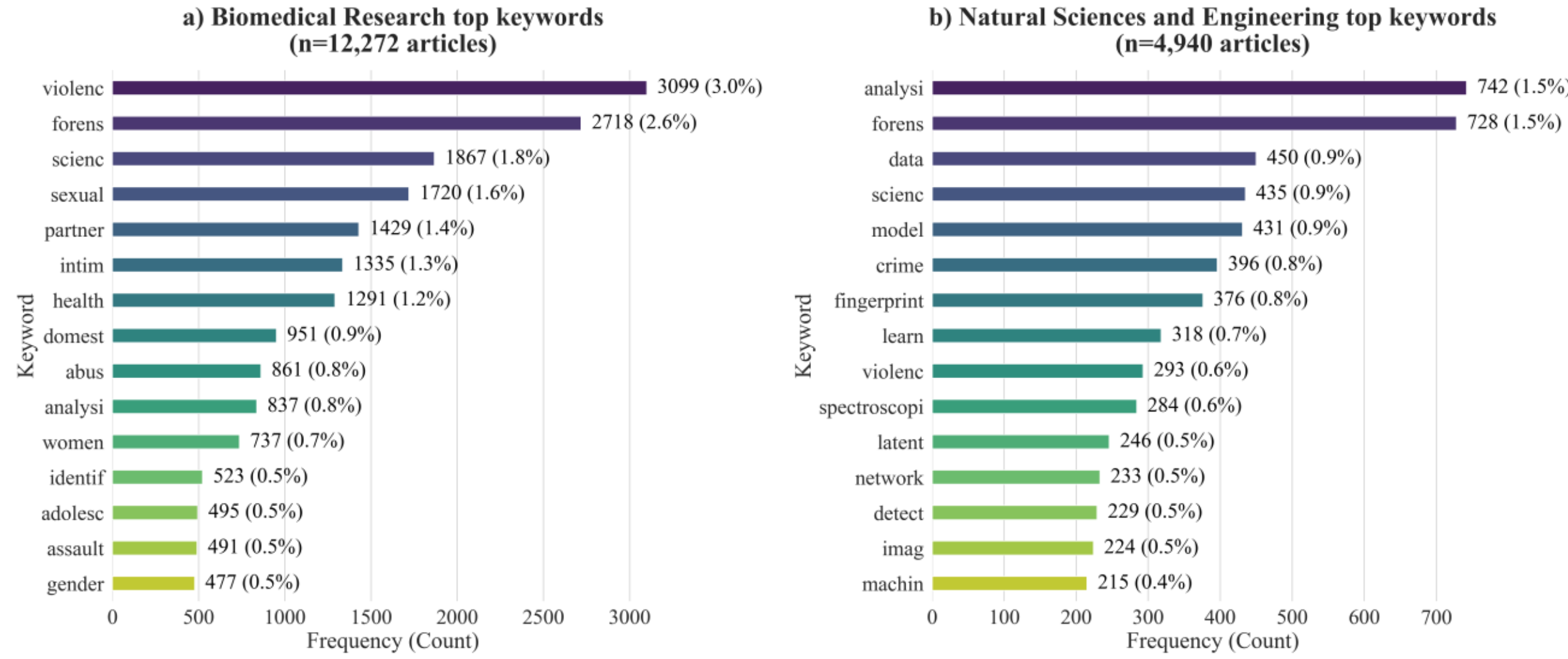

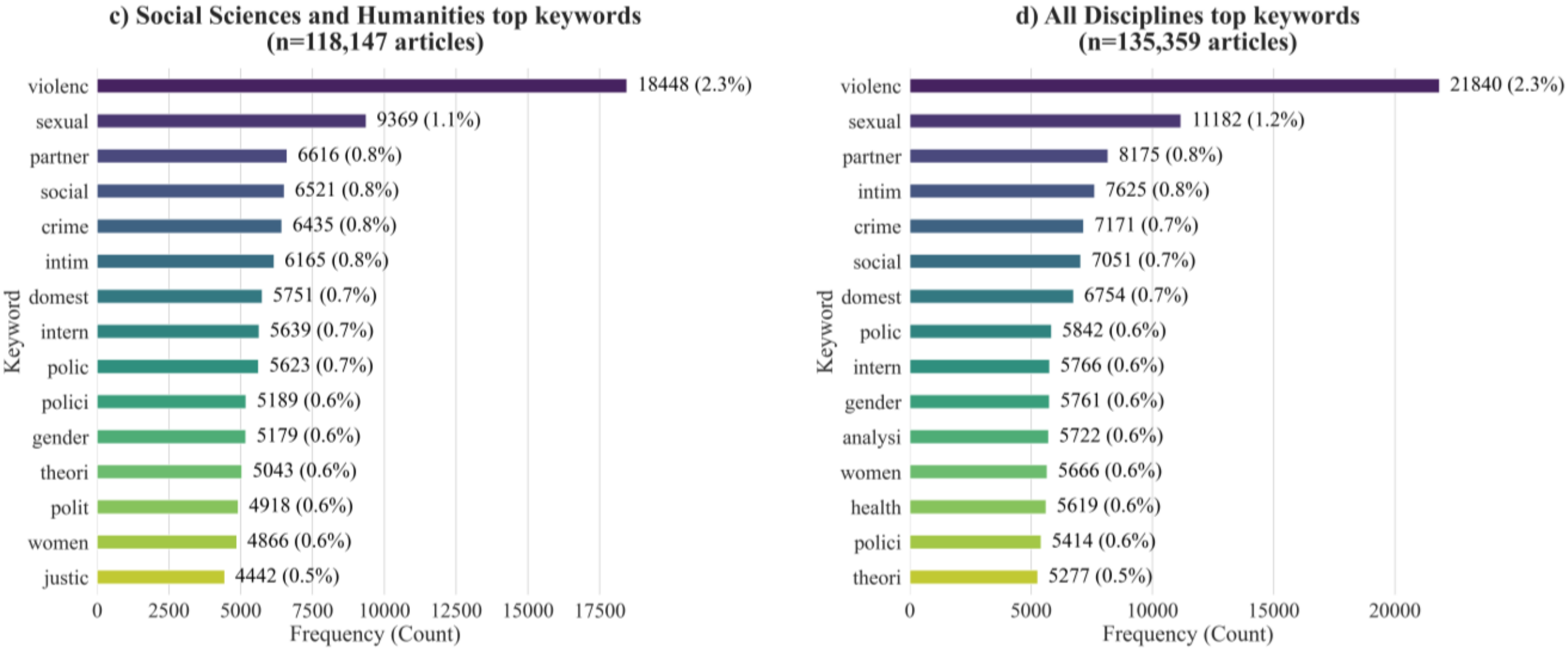

*Figure 2 - Top keywords in peace-related papers by discipline*

Figures 3 and 4 show that peace is characterized differently across countries with high and low levels of peacefulness.

**GPI_2. New Zealand (n=283 papers)**
violenc 54 (3.2%); sexual 23 (1.3%); australia 22 (1.3%); partner 22 (1.3%); offend 21 (1.2%); crime 20 (1.2%); health 18 (1.1%); intim 18 (1.1%); polici 18 (1.1%); risk 18 (1.1%)
0 10 20 30 40 50
Frequency

**GPI_3. Switzerland (n=135 papers)**
sexual 14 (1.7%); violenc 12 (1.5%); prison 11 (1.4%); health 10 (1.2%); polici 10 (1.2%); crime 7 (0.9%); foreign 7 (0.9%); intern 7 (0.9%); right 7 (0.9%); social 7 (0.9%)
0 2 4 6 8 10 12 14
Frequency

**GPI_4. Slovenia (n=197 papers)**
crime 86 (6.6%); polic 58 (4.5%); prison 23 (1.8%); investig 21 (1.6%); offenc 21 (1.6%); legitimaci 19 (1.5%); commun 17 (1.3%); secur 16 (1.2%); violenc 15 (1.2%); offic 13 (1.0%)
0 20 40 60 80
Frequency

**GPI_5. Ireland (n=536 papers)**
violenc 51 (2.1%); conflict 35 (1.5%); sexual 32 (1.3%); polic 30 (1.3%); polit 30 (1.3%); terror 28 (1.2%); justic 26 (1.1%); prison 26 (1.1%); memori 23 (1.0%); peac 20 (0.8%)
0 10 20 30 40 50
Frequency

**GPI_6. Austria (n=125 papers)**
memori 11 (1.8%); nation 9 (1.5%); polici 8 (1.3%); polit 8 (1.3%); collect 7 (1.1%); foreign 7 (1.1%); crime 6 (1.0%); offend 6 (1.0%); secur 6 (1.0%); social 6 (1.0%)
0 2 4 6 8 10
Frequency

**GPI_7. Portugal (n=197 papers)**
violenc 50 (3.6%); partner 26 (1.9%); intim 24 (1.7%); victim 21 (1.5%); sexual 19 (1.4%); women 18 (1.3%); abus 16 (1.2%); domest 14 (1.0%); gender 13 (0.9%); justic 13 (0.9%)
0 10 20 30 40 50
Frequency

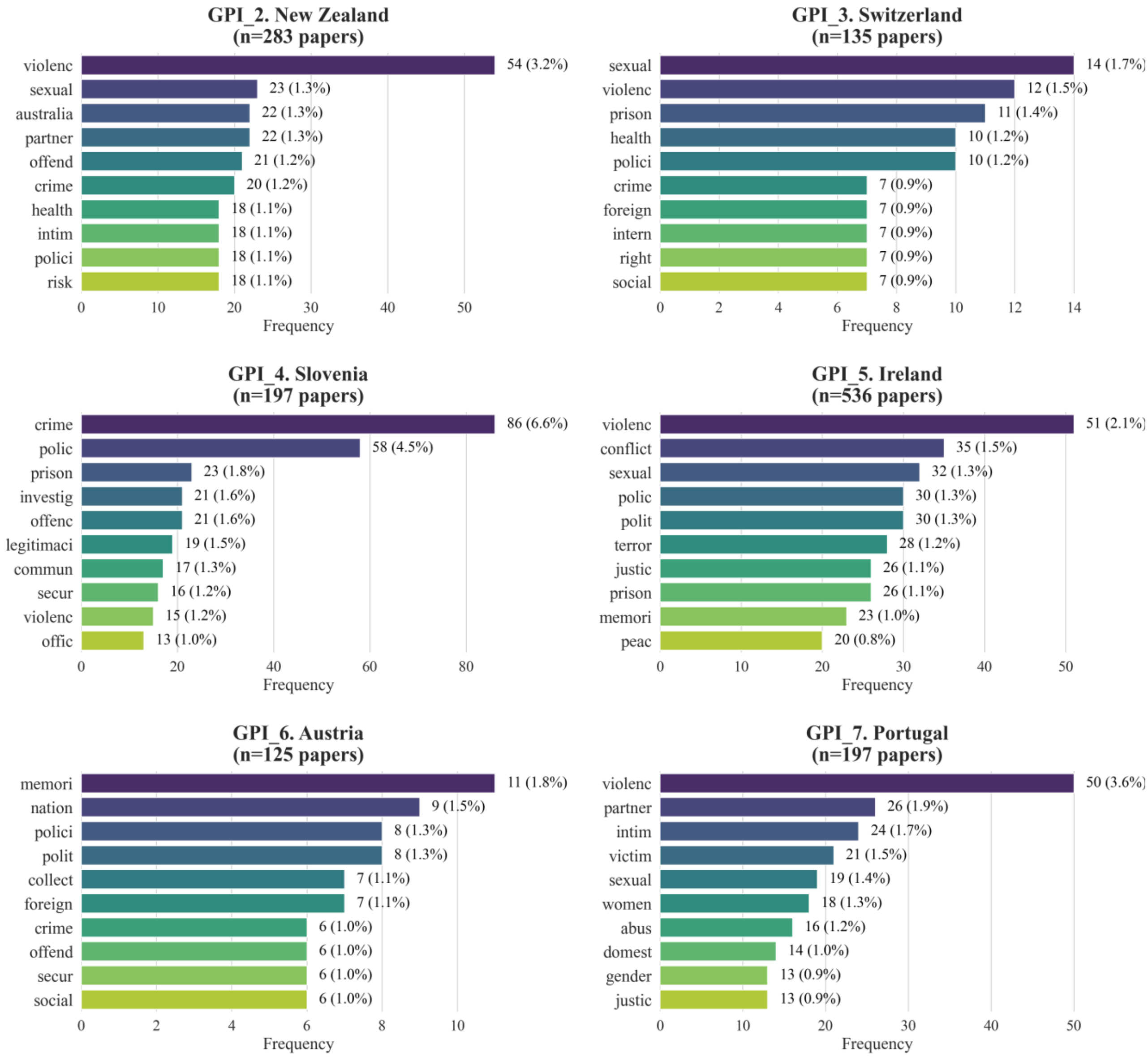

*Figure 3 - Top keywords in peace-related papers from the most peaceful countries. GPI# indicates the country's GPI rank.*

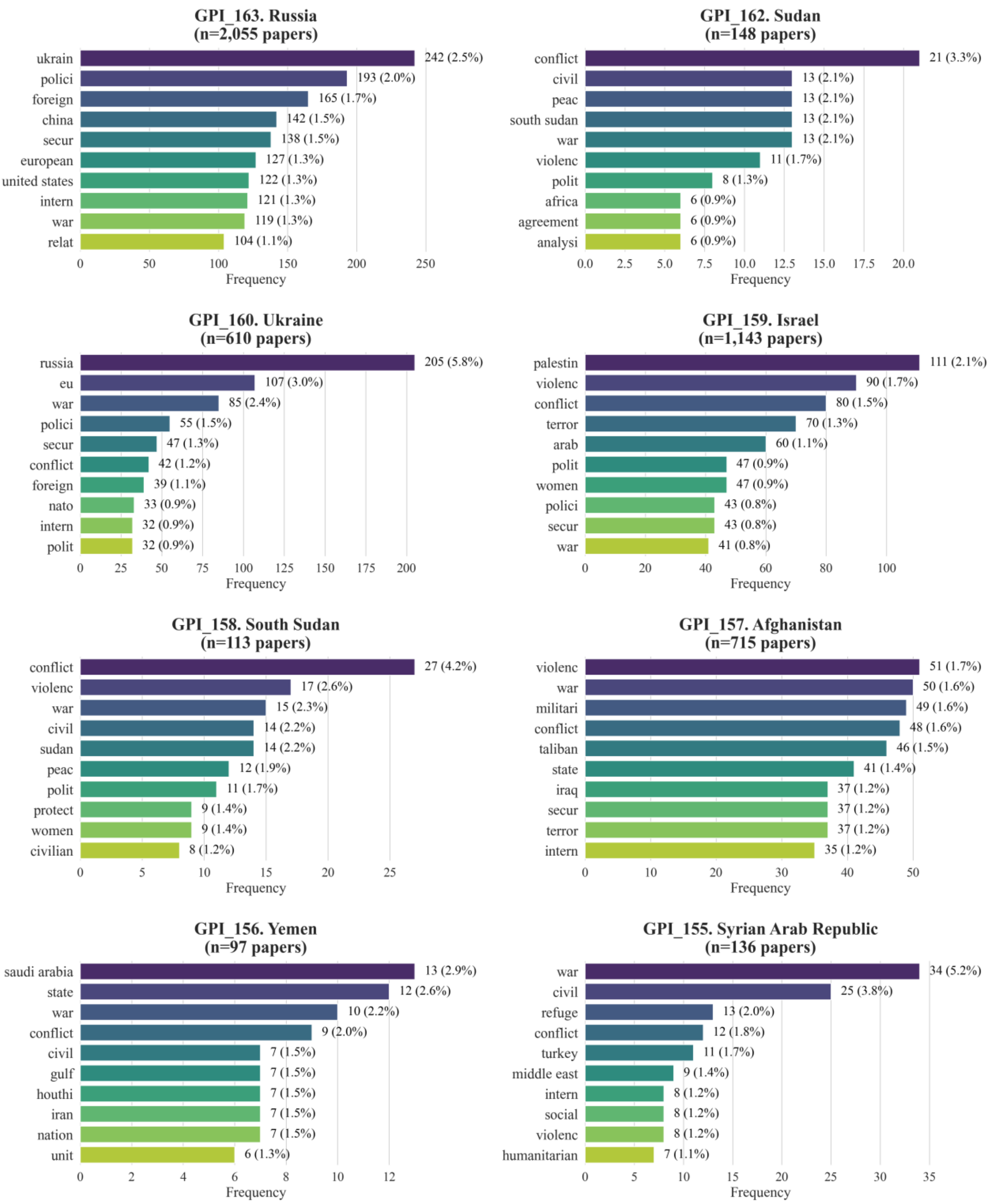

*Figure 4 - Top keywords in peace-related papers from the least peaceful countries. GPI# indicates the country's GPI rank.*

## 4. Discussions and Conclusion

This paper in progress analyzes how peace is characterized by countries in scientific papers and identifies which keywords are most strongly associated with peace-related publications. Overall, the results suggest that peace-related research is strongly centred on violence—particularly sexual, domestic, and intimate partner violence—examined through both forensic/technological and social/institutional lenses. While biomedical and NSE research emphasize detection, analysis, and technological tools, the social sciences focus more on governance, social justice, and gendered power structures.

At the country level, violence remains a central anchor across all contexts, but its framing varies systematically with levels of peace. In higher-peace countries, violence is more often embedded within welfare-oriented and institutional frameworks, with greater attention to health, gender equality, family, and social support systems. In lower-peace contexts, the discourse shifts toward conflict, security, and geopolitics, with strong associations to war, security and international relations. Taken together, these patterns suggest that peace is not conceptualized as the absence of violence, but rather as a condition shaped by how violence is understood, managed, and situated within broader geopolitical systems.
These findings are important because they show that peace is constructed differently across disciplines and geopolitical contexts, shaping both research priorities and policy initiatives. Moving forward, this work will be extended by developing interactive maps to visualize country-level keyword clusters, conducting network analyses of keyword co-occurrences, and incorporating qualitative discourse analysis. In addition, future analysis of abstracts will examine the extent to which peace is framed as promoting justice and human rights, rather than merely as the absence of conflict.

**Author contribution**

**Gita Ghiasi.**: Conceptualization, Methodology, Investigation, Data Curation, Supervision, Writing- Original draft preparation. **Rabeeh Parhizkari:** Methodology, Visualization, Investigation, Data curation. **Tanja Tajmel:** Conceptualization

**References**

Armitage, C. S., Lorenz, M., & Mikki, S. (2020). Mapping scholarly publications related to the Sustainable Development Goals: Do independent bibliometric approaches get the same results? *Quantitative Science Studies*, *1*(3), 1092–1108. https://doi.org/10.1162/qss_a_00071

Coleman, P. T., Fisher, J., Fry, D. P., Liebovitch, L. S., Chen-Carrel, A., & Souillac, G. (2021). How to live in peace? Mapping the science of sustaining peace: A progress report. *American Psychologist*, *76*(7), 1113–1127. https://doi.org/10.1037/amp0000745

Diehl, P. F. (2016). Exploring Peace: Looking Beyond War and Negative Peace. *International Studies Quarterly*, *60*(1), 1–10. https://doi.org/10.1093/isq/sqw005

Fry, D. P., Souillac, G., Liebovitch, L., Coleman, P. T., Agan, K., Nicholson-Cox, E., Mason, D., Gomez, F. P., & Strauss, S. (2021). Societies within peace systems avoid war and build positive intergroup relationships. *Humanities and Social Sciences Communications*, *8*(1), 17. https://doi.org/10.1057/s41599-020-00692-8

Garcia, M. (2022, February 3). *A more sustainable future for all: Introducing the UN Sustainable Development Goals in InCites | Clarivate*. https://clarivate.com/academia-government/blog/a-more-sustainable-future-for-all-introducing-the-un-sustainable-development-goals-in-incites/

Ghiasi, G., Harsh, M., Tajmel, T., & Larivière, V. (2021). Where international development and gender equality meet in science: A bibliometric analysis. *Proceedings of the 18th International Conference on Scientometrics & Informetrics*, 447–452. https://www.ost.uqam.ca/publications/where-international-development-and-gender-equality-meet-in-science-a-bibliometric-analysis/

Institute for Economics & Peace. (2026). *Global Peace Index 2026*. Institute for Economics & Peace.

Liebovitch, L. S., Powers, W., Shi, L., Chen-Carrel, A., Loustaunau, P., & Coleman, P. T. (2023). Word differences in news media of lower and higher peace countries revealed by natural language processing and machine learning. *PLOS ONE*, *18*(11), e0292604. https://doi.org/10.1371/journal.pone.0292604

Mahmoud, Y., & Makoond, A. (2017). *Sustaining Peace: What Does It Mean in Practice?* International Peace Institute. https://www.ipinst.org/2017/04/sustaining-peace-in-practice

Prasad, T., Liebovitch, L. S., Wild, M., West, H., & Coleman, P. T. (2025). Words that Represent Peace. *2025 59th Annual Conference on Information Sciences and Systems (CISS)*, 1–6. https://doi.org/10.1109/CISS64860.2025.10944736

United Nations. (2015). *Transforming our World: The 2030 Agenda for Sustainable Development*. https://sdgs.un.org/sites/default/files/publications/21252030%20Agenda%20for%20Sustainable%20Development%20web.pdf